\documentstyle[twocolumn,prl,aps,epsf,floats]{revtex}
\begin{document}
\newcommand{\so}
   {\mathrel{\rlap{\raise1pt\hbox{$>$}}{\lower4pt\hbox{$\sim$}}}}
\draft

\twocolumn[\hsize\textwidth\columnwidth\hsize\csname@twocolumnfalse\endcsname

\title{Coarse-grained surface energies and temperature-induced\\ anchoring
transitions in nematic liquid crystals}

\author{J.-B. Fournier and P. Galatola~\cite{turin}}
\address{Laboratoire de Physico-Chimie Th\'eorique, E.\,S.\,P.\,C.\,I.,
10 rue Vauquelin, F-75231 Paris C\'edex 05, France}
\date{\today}
\maketitle
\begin{abstract}
  We introduce a coarse-grained description of the surface energy of a
  nematic liquid crystal. The thermal fluctuations of the nematic
  director close to the surface renormalize at macroscopic scales the
  bare surface potential in a temperature-dependent way. The angular
  dependence of the renormalized potential is dramatically smoothed,
  thus explaining the success of the Rapini-Papoular form. Close
  to the isotropic phase, the anchoring energy is strongly suppressed
  and the change of its shape allows for anchoring transitions. Our
  theory describes quantitatively the temperature dependence of the
  anchoring energy and the temperature-induced anchoring transitions
  reported in the literature.
\end{abstract}
\twocolumn\vskip.5pc]\narrowtext

In recent years, surface phenomena have attracted a lot of interest.
Particularly, the interface between liquid crystals~\cite{deGennes}
and solid substrates displays a rich variety of behaviors:
orientational wetting~\cite{Sheng,Miyano,Lipowsky,Crawford} and
spreading~\cite{Xue,Valignat}, memory effects~\cite{Ouchi}, surface
melting~\cite{Barberi}, Kosterlitz-Thouless
transitions~\cite{Sluckin}, quasi-critical behavior of surface
energies~\cite{Faetti,Seo,Peng}, surface
anchoring transitions~\cite{Jerome_rev,Kleman,Monkade,Jerome,Bechhoefer,%
  Teixeira,Patel,Zhu,Pontus}, etc. Despite the complexity and the
diversity of the interactions between nematic liquid crystals and
solid substrates, some simple---though unexplained---ubiquitous
aspects emerge from the experiments: the angular dependences of the
surface potentials are extremely smooth and well described by the
so-called Rapini-Papoular law~\cite{Rapini}, at high temperatures the
preferred surface orientation is usually either parallel or
perpendicular to the substrate, tilted orientations are difficult to
achieve~\cite{Filas}, and temperature-driven anchoring transitions
systematically occur immediately below the bulk isotropic transition
temperature~\cite{Patel}. In this Letter, we show that all these
effects can be explained in terms of a renormalization of the surface
energy by the short-wavelength orientational fluctuations in the bulk.

Nematic liquid crystals are fluid mesophases made of elongated
molecules displaying a broken orientational symmetry along a non-polar
direction called the nematic director ${\bf n}$~\cite{deGennes}. At a
molecular level, nematics exhibit large orientational fluctuations and
usually some degree of short-range biaxial and positional order.  The
link between the microscopic and the macroscopic description can be
established by means of a coarse-graining procedure~\cite{Chaikin}.
Before determining the consequences of this procedure on the surface
behavior, let us briefly discuss how it is carried out in the bulk.
One can associate a local orientation ${\bf m}$ to each molecule; then
the probability $p[{\bf m}]$ of a given instantaneous microscopic
configuration ${\bf m}({\bf r})$ is proportional to the Boltzmann
factor $\exp(-\beta{\cal F}[{\bf m}])$, where ${\cal F}[{\bf m}]$ is a
complex microscopic free-energy~\cite{note} and $\beta=1/k_{\rm B}T$
is the inverse temperature.  At macroscopic scales the bulk
free-energy $F[{\bf n}]$ is well described by a simple elasticity
involving only the average molecular orientation ${\bf n}$ and its
gradient~\cite{Frank,deGennes}. The quantity $\exp(-\beta F[{\bf n}])$
gives the probability of observing a given smooth director
configuration ${\bf n}({\bf r})$, whatever the details of its rapidly
varying components. One therefore obtains the coarse-grained
free-energy $F[{\bf n}]$ by summing the microscopic probabilities
$\exp(-\beta{\cal F}[{\bf m}])$ over all the Fourier components ${\bf
m}({\bf q})$ with wavevector $|{\bf q}|>\Lambda$, where $\Lambda$ is
some macroscopic cutoff, the components ${\bf m}({\bf q})$ with $|{\bf
q}|<\Lambda$ being fixed and denoted by ${\bf n}({\bf q})$. The resulting
free-energy $F[{\bf n}]$, which corresponds to the Frank
elasticity~\cite{Frank}, is therefore meaningful only for the slowly
varying Fourier components $|{\bf q}|<\Lambda$ of ${\bf n}$. Such a
coarse-graining procedure yields a renormalized elasticity expanded in
power series of the derivatives of ${\bf n}$, whose coefficients
depend on $T$ and are expected to scale as powers of the range $b$ of
the molecular interactions. This is why, for $\Lambda b\ll1$, it is
justified to retain in $F[{\bf n}]$ only the lowest-order gradient
terms.

Using the Frank elasticity $F[{\bf n}]$ implicitly entails a
coarse-graining in the bulk.  For the sake of consistency, this
procedure should also be performed on the surface. The latter will
then be effectively transformed into a blurred layer of width
$\Lambda^{-1}$ acquiring some properties of the bulk. Such a coupling
between surface and bulk is usually introduced in a phenomenological
way by means of Landau expansions in the tensorial nematic
order-parameter~${\bf
  Q}$~\cite{Sheng,Miyano,Lipowsky,Barberi,Sluckin,Faetti,Seo,Peng,%
  Jerome_rev,Barbero,Sen}. However, modeling surface properties in
this way is rather complicated, since one has to deal with spatially
varying tensorial fields, and also somewhat arbitrary, since high
powers of ${\bf Q}$ should be included due to the first-order
character of the nematic--isotropic transition.

\begin{figure}
\centerline{\hspace{0cm}\epsfxsize=8.5cm\epsfbox{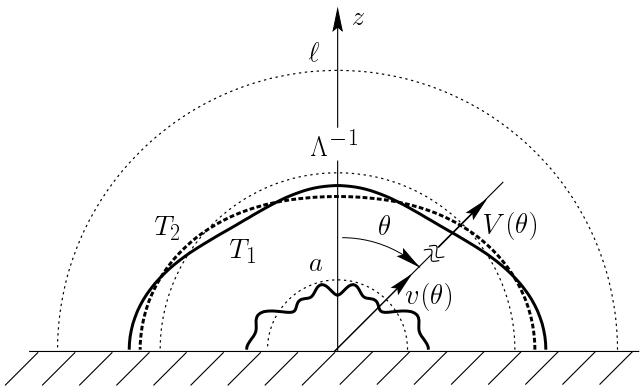}\hspace{0cm}}
  \caption{Schematic representation of the surface and its anchoring
    potential at the mesoscopic scale $a$ and coarse-grained scale
    $\Lambda^{-1}$. The mesoscopic potential $v(\theta)$ and its
    corresponding effective coarse-grained potential $V(\theta)$ are
    plotted in polar coordinates in arbitrary units.  The renormalized
    potential $V(\theta)$, which depends on temperature, favors here
    an oblique anchoring at $T_1$ and an homeotropic one at
    $T_2>T_1$.}
  \label{schema}
\end{figure}

We have calculated the renormalization of the surface energy by
coarse-graining the director orientation from a mesoscopic length $a$,
of the order of the nematic coherence length $\xi_{\rm NI}$ at the
isotropic transition~\cite{deGennes}, up to a scale
$\Lambda^{-1}$ at least a few times larger than $a$. Our results,
based on perturbation theory, are valid for weak anchorings, in the
sense $\ell\gg\xi_{\rm NI}$, where $\ell$ is the anchoring
extrapolation length~\cite{deGennes}.  Calling $\theta$ and $\phi$ the
director's polar angles and starting from a ``bare'' surface anchoring
energy expanded in Fourier harmonics of the form ${\rm
  cs}(2n\theta)\,{\rm cs}(m\phi)$ (${\rm cs}$ being either $\cos$ or
$\sin$), we find that each harmonic is independently renormalized by a
Debye-Waller factor $\exp(-\alpha_n\!-\!\beta_m)$, with
$\alpha_n,\beta_n\propto n^2\,k_{\rm B}T/Ka$, where $K$ is a bulk
nematic elastic constant. At scales larger than $\Lambda^{-1}$, the
high-order harmonics are thus strongly suppressed by the director
thermal fluctuations: this explains the success of the Rapini-Papoular
form~\cite{Rapini}.  Moreover, the anchoring energy naturally acquires
a temperature dependence through the elastic constant $K(T)$. The
surface energy is thus reduced close to the nematic--isotropic
transition---where $K$ is lowered---in agreement with
experiments~\cite{Faetti,Seo,Peng}. The different temperature
dependences of the surface harmonics allow for anchoring transitions:
as the temperature increases, the suppression of the high-order
harmonics shifts the minimum of the anchoring energy towards some
symmetry axis of the surface. Our results fit well the quasi-critical
temperature dependence of the azimuthal anchoring energy measured by
Faetti et al.~\cite{Faetti}, the oblique-to-homeotropic anchoring
transition measured by Patel and Yokoyama~\cite{Patel}, and the
bistable oblique-to-planar anchoring transition~\cite{Monkade}
measured by J\"agemalm et al.~\cite{Pontus}.  It turns out that the
effective macroscopic anchoring, whose minimum gives the
coarse-grained director orientation at the surface, can significantly
differ from the mesoscopic surface potential (Fig.~\ref{schema}). The
director fluctuations within the coarse-grained region dramatically
smooth the fine details of the anchoring energy. They can also shift
the average equilibrium position of the director at the surface,
similarly to the amplitude-dependent shift of the average position of
an anharmonic oscillator~\cite{Landau}.

Precisely, we consider a semi-infinite nematic slab in the $z\ge0$
half-space and we describe the nematic director by its spherical
coordinates $\theta$, $\phi$ centered on the $z$-axis. We start with a
bulk elasticity already coarse-grained on a mesoscopic length
$a\approx\xi_{\rm NI}$, such that it can be expressed in the usual
Frank form. In the one-constant approximation~\cite{deGennes}, its
harmonic part, expanded about an arbitrary direction ($\theta_0$,
$\phi_0$), takes the form
\begin{equation}
{\cal H}_0=\frac{1}{2}K(T)\int\!d^3r\left[
\left(\nabla\theta\right)^2+
\sin^2\theta_0\left(\nabla\phi\right)^2
\right].
\end{equation}
At the length-scale $a$, the ``bare'' surface potential ${\cal H}_s$
is given by some local functional of the mesoscopic director's
orientation at the surface
\begin{equation}
{\cal H}_s=\int\!d^2r_\perp\,v(\theta,\phi).
\end{equation}
The total free-energy $F_t$ of the nematic slab is given by the
path integral
\begin{equation}
\exp(-\beta F_t)=
\int\!\!{\cal D}[\theta]{\cal D}[\phi]\,\exp\left[-\beta(
{\cal H}_0+{\cal H}_{\rm nh}+{\cal H}_s)\right],
\end{equation}
where ${\cal H}_{\rm nh}$ contains the non-harmonic bulk terms. To
further coarse-grain on a length-scale $\Lambda^{-1}\!>\!a$, we put
$\theta({\bf r})=\theta^<({\bf r})+\theta^>({\bf r})$ and $\phi({\bf
  r})=\phi^<({\bf r})+\phi^>({\bf r})$, where $\theta^<({\bf r})$,
$\phi^<({\bf r})$ have Fourier components with wavevectors $|{\bf
  q}|\le\Lambda$, and $\theta^>({\bf r})$, $\phi^>({\bf r})$ have
Fourier components with $\Lambda<|{\bf q}|<2\pi/a$. Integrating out
$\theta^>$ and $\phi^>$ yields the renormalized Hamiltonian ${\cal
  H}^<$ at length-scales $\Lambda^{-1}$; to lowest-order in
perturbation theory, it is given by~\cite{Chaikin}
\begin{equation}\label{Hinf}
{\cal H}^<={\cal H}_0+\langle{\cal H}_{\rm nh}+{\cal H}_s\rangle_>
\end{equation}
where $\langle\cdots\rangle_>$ indicates the statistical average over
the high-wavevector components, weighted by the Gaussian Hamiltonian
${\cal H}_0$. At first-order, the bulk and the surface are therefore
renormalized independently.

Due to the ${\bf n}\to-{\bf n}$ invariance, the bare surface energy
density can be Fourier expanded as
\begin{eqnarray}
v(\theta,\phi)={\sum_{n,m}}'
\cos n\theta&&\big[w_{nm}^{\rm cc}\,\cos m\phi
+w_{nm}^{\rm cs}\,\sin m\phi\big]\nonumber\\
+{\sum_{n,m}}''
\sin n\theta&&\big[w_{nm}^{\rm sc}\,\cos m\phi
+w_{nm}^{\rm ss}\,\sin m\phi\big],
\label{Fourier}\end{eqnarray}
where $n$ runs over all even integers, and $m$ runs over all
even (resp.\ odd) integers in the first (resp.\ second) sum.

Even though~(\ref{Fourier}) is non-linear, the renormalized
Hamiltonian~(\ref{Hinf}) can be transformed into Gaussian integrals by
writing the trigonometric functions as complex exponentials. We obtain
a renormalized surface energy density $V(\theta^<,\phi^<)$ whose
Fourier components $W_{nm}^{\alpha\beta}$ ($\alpha$, $\beta$ being
either c or~s) are separately renormalized:
\begin{equation}\label{ren}
W_{nm}^{\alpha\beta}(t)=w_{nm}^{\alpha\beta}\exp\left[-\frac{1}{2}\left(
n^2+\frac{m^2}{\sin^2\theta_0}\right)t\right],
\end{equation}
where, for $\Lambda a\ll1$,
\begin{equation}
t=\langle\theta^>({\bf r})^2\rangle_>=
k_{\rm B}T\!\int_\Lambda^{2\pi/a}\!\!\frac{d^3q}
{(2\pi)^3}\frac{1}{Kq^2}\simeq
\frac{k_{\rm B}T}{\pi Ka}.
\end{equation}
Note that $\langle\phi^>({\bf
  r})^2\rangle_>=t/\sin^2 \theta_0$ also appears in~(\ref{ren}).

\begin{figure}
\centerline{\hspace{0cm}\epsfxsize=8cm\epsfbox{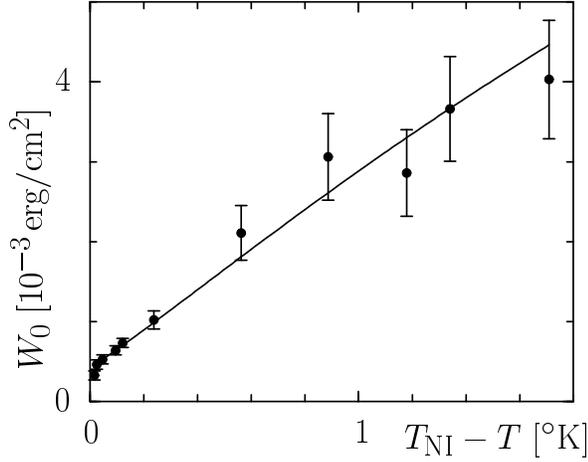}\hspace{0cm}}
  \caption{Azimuthal anchoring energy $W_0$ vs.\ temperature $T$
    according to Faetti et al.~\protect\cite{Faetti}. Points:
    experimental data; solid line: our fit.}
  \label{w0}
\end{figure}
It follows from~(\ref{ren}) that for large $t$, i.e., in the vicinity
of the nematic-isotropic transition, where $K$ is reduced, the
effective surface potential should be very-well described by its first
harmonics. This is indeed what was measured by Faetti et
al.~\cite{Faetti} using the nematic liquid crystal 5CB, planarly
anchored on a glass plate treated by oblique evaporation of SiO:
exploring the azimuthal anchoring energy in regions far away from the
parabolic minimum, they could not detect any deviation from the simple
form
\begin{equation}
V=W_0\,\sin^2\phi=W_{02}^{\rm cc} \cos2\phi+{\rm const}.
\end{equation}
Setting $\theta_0=\pi/2$, $\phi_0=0$, we have fitted their data
with~(\ref{ren}), i.e., $W_{02}^{\rm cc}(t)=w_{02}^{\rm cc}\exp(-2t)$, in
which we have assumed a Landau-de Gennes form for the temperature
dependence of $K$~\cite{Ping}, yielding
\begin{equation}\label{t}
t=\frac{k_{\rm B}T_{\rm NI}}{\pi K_{\rm NI}\,a}
\left(\frac{4}{3+\sqrt{9-8\tau}}
\right)^2,
\end{equation}
where $\tau=(T-T^*)/(T_{\rm NI}-T^*)$ is the reduced temperature,
$T_{\rm NI}$ the nematic-isotropic transition temperature, $T^*$ the
isotropic supercooling temperature, and $K_{\rm NI}$ the elastic
constant at the transition. With $T_{\rm NI}\simeq306.8\,^\circ{\rm
  K}$, $T_{\rm NI}-T^*\simeq1.5\,^\circ{\rm K}$~\cite{Ioannis}, our
best fit, shown in Fig.~\ref{w0}, yields $t_{\rm NI}\equiv k_{\rm
  B}T_{\rm NI}/\pi K_{\rm NI}\,a=2.1\pm0.1$ and $W_{02}^{\rm cc}(t_{\rm
  NI})=(-2.08\pm0.08)\times10^{-4}\,{\rm erg/cm}^2$. For an
extrapolated $K_{\rm NI}\simeq0.5\,{\rm pN}$~\cite{Bunning}, the value
of $t_{\rm NI}$ yields $a\simeq13\,{\rm\AA}$, which roughly compares
with $\xi_{\rm NI}$ for a first-order phase transition.

\begin{figure}
\centerline{\hspace{0cm}\epsfxsize=8.2cm\epsfbox{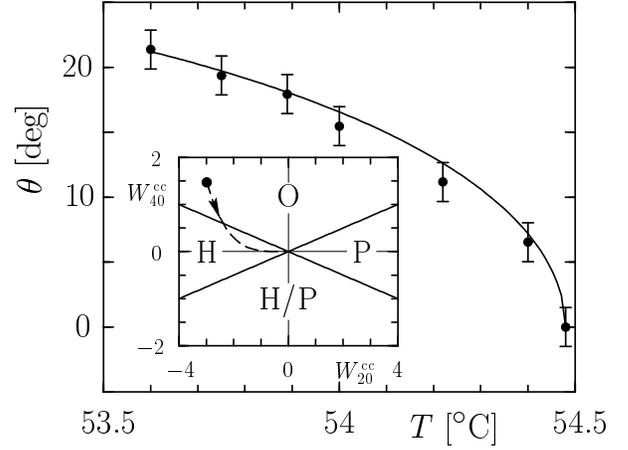}\hspace{0cm}}
  \caption{Surface easy-axis $\theta$ (measured with respect to the
    surface normal) vs.\ temperature $T$ according to Patel and
    Yokoyama~\protect\cite{Patel}. Points: experimental data; solid
    line: our fit. Inset: anchoring phase diagram (in arbitrary units)
    corresponding to Eq.~(\protect\ref{oh}).}
  \label{patel}
\end{figure}

Close to the nematic-isotropic transition of the nematic compound E7,
Patel and Yokoyama~\cite{Patel} have observed a tilted-to-homeotropic
anchoring transition on a fluoropolymer-coated surface. Again, we can
explain it by retaining in $V$ the first two harmonics
allowed by symmetry, i.e.,
\begin{equation}\label{oh}
V=W_{20}^{\rm cc} \cos2\theta+W_{40}^{\rm cc} \cos4\theta.
\end{equation}
Minimizing~(\ref{oh}) with respect to $\theta$ yields the anchoring
phase diagram shown in the inset of Fig.~\ref{patel}, which displays
the following regions separated by second-order transition lines:
conical oblique (O), homeotropic (H), degenerated planar (P), and
metastable homeotropic/planar (H/P). According to (\ref{ren})
and~(\ref{t}), as temperature increases, the easy-axis of the
anchoring follows the typical path shown in the phase diagram. Setting
$T_{\rm NI}\simeq57.7\,^\circ{\rm C}$ and $T_{\rm H}=54.48\,^\circ{\rm
  C}$~\cite{Patel}, where $T_{\rm H}$ is the homeotropic-to-oblique
transition temperature, we fit closely the temperature dependence of
the easy-axis direction $\theta$ with $T_{\rm
  NI}-T^*\simeq1.62\,^\circ{\rm K}$ and $t_{\rm NI}=1.2\pm0.04$
(Fig.~\ref{patel}).

Using the same values of $T_{\rm NI}$, $T_{\rm NI}-T^*$ and $t_{\rm
  NI}$, which depend only on the nematic material, we can fit equally
well the bistable oblique-to-planar anchoring transition observed by
J\"agemalm et al.~\cite{Pontus} at $T=T_{\rm P}=49.7\,^\circ{\rm C}$
for the same compound E7 (Fig.~\ref{pontus}). This anchoring
transition also appears close to the nematic-isotropic transition, in
the presence of a glass substrate treated by an oblique evaporation of
SiO. Here, by symmetry, the lowest-order expansion of $V$ is
\begin{equation}\label{bp}
V=W_{20}^{\rm cc} \cos2\theta+W_{02}^{\rm cc} \cos2\phi+
W_{21}^{\rm sc} \sin2\theta\,\cos\phi.
\end{equation}
Setting $\overline{W}_\theta=W_{20}^{\rm cc}/W_{02}^{\rm cc}$ and
$\overline{W}_{\theta\phi}=|W_{21}^{\rm sc}/W_{02}^{\rm cc}|$, the
corresponding anchoring phase diagram, shown in the inset of
Fig.~\ref{pontus}, displays the four regions, separated by
second-order transition lines, that are observed
experimentally~\cite{Monkade,Jerome}: oblique in the evaporation plane
(O), bistable symmetric with respect to the evaporation plane (B),
homeotropic (H), and planar orthogonal to the evaporation plane
(P). Once fixed the values of $T_{\rm P}$, and, by the
previous fit, of $T_{\rm NI}$, $T_{\rm NI}-T^*$, and $t_{\rm NI}$, the
fit on $\theta$ has no adjustable parameter, while for the fit on
$\phi$ there remains the only free parameter $w_{21}^{\rm
sc}/w_{02}^{\rm cc}$, best adjusted to the value $6.76\pm0.05$.

\begin{figure}
\centerline{\hspace{0cm}\epsfxsize=8cm\epsfbox{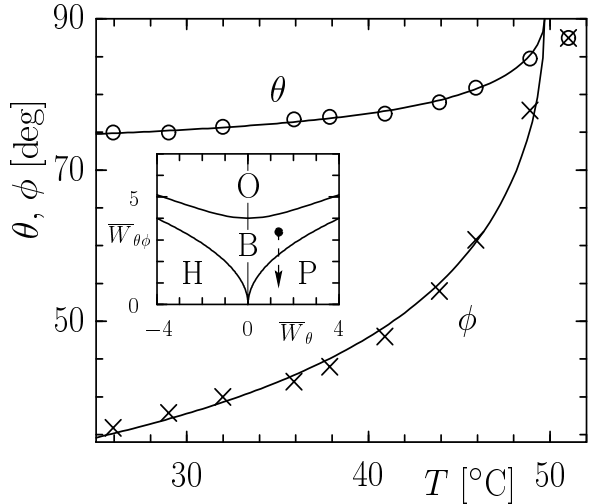}\hspace{0cm}}
  \caption{Surface easy-axis ($\theta$, $\phi$)
    vs.\ temperature $T$ according to J\"agemalm et
    al.~\protect\cite{Pontus}.  Points: experimental data; solid line:
    our fit. Inset: anchoring phase diagram corresponding to
    Eq.~(\protect\ref{bp}); the arrow indicates the path followed as
    $T$ increases.}
  \label{pontus}
\end{figure}

To estimate the validity of our first-order perturbative expansion in
the surface potential, we have calculated the second-order correction
to $W_{40}^{\rm cc}$ coming from $w_{20}^{\rm cc}$~\cite{note2}. For
$\Lambda a\ll1$, we find
\begin{equation}
\frac{\delta W_{40}^{\rm cc}}{w_{20}^{\rm cc}}\simeq
\frac{2\,w_{20}^{\rm cc}}{\pi\Lambda K}\,\exp(-4t)\,,
\end{equation}
which is negligible for $\Lambda^{-1}\ll\ell$, where
$\ell=K/w_{20}^{\rm cc}$ is a bare anchoring extrapolation length;
this sets the limit of validity of our analysis.  Coarse-graining on a
length $\Lambda^{-1}\so\ell$ effectively reduces the anchoring for
purely elastic reasons, as will be described elsewhere.

Finally, note that, since we coarse-grained on a length
$\Lambda^{-1}>\xi_{\rm NI}$, our model is insensitive to the
variations of the scalar order-parameter $S$~\cite{deGennes}. This
does not imply, however, that $S$ must be uniform throughout the
sample: any substrate inducing a surface variation of $S$ will still
be described---at length-scales larger than $\xi_{\rm NI}$---by some
potential $v(\theta,\phi)$ depending only on the director's
orientation. Our analysis simply disregards the underlying surface
variations of $S$: this is a limitation only when $\xi_{\rm NI}$
becomes critically large.

We thank A. Ajdari, G. Durand, P.-G. de Gennes, and L. Peliti for
useful discussions. P. G. acknowledges the support of a
Chaire Joliot de l'ESPCI.


\begin{references}
\bibitem{note2}Actually, all orders can be calculated separately by
  means of Gaussian integrals.
\end{references}
\end{document}